\documentclass[aps,pra,preprint,showpacs,amsmath,nofootinbib,superscriptaddress]{revtex4}

\usepackage{graphicx}

\begin{document}

\title{Embedding quantum systems with a non-conserved probability
in classical environments}

\author{Alessandro Sergi
}
\email{sergi@ukzn.ac.za}

\affiliation{
School of Chemistry and Physics, University of KwaZulu-Natal in Pietermaritzburg,
Private Bag X01, Scottsville 3209, South Africa}
\affiliation{
KwaZulu-Natal Node,
National Institute for Theoretical Physics (NITheP), South Africa}

\begin{abstract}
Quantum systems with a non-conserved probability
can be described by means of
non-Hermitian Hamiltonians and non-unitary dynamics.
In this paper, the case in which
the degrees of freedom can be partitioned in two subsets 
with light and heavy masses is treated.
A classical limit over the heavy coordinates is taken
in order to embed the non-unitary dynamics of the subsystem
in a classical environment. Such a classical environment, in turn, acts
as an additional source of dissipation (or noise),
beyond that represented by the non-unitary evolution.
The non-Hermitian dynamics of a Heisenberg two-spin chain, with the spins
independently coupled to 
harmonic oscillators, is considered in order to illustrate the formalism.
\end{abstract}

\maketitle

\section{Introduction}

Historically, the development of non-Hermitian quantum mechanics
started with the study of metastable states
and tunneling by means of Hamiltonians with
complex energy eigenvalues~\cite{gamow,landau}.
Later a connection to ${\cal PT}$-symmetry was found~\cite{bender}
and the concept of pseudo-hermiticity was
also established~\cite{mostafazadeh,mostafazadeh2}.
Nowadays, this field of research is constantly growing.
A first general book on the topic has appeared~\cite{moiseyev};
applications of non-Hermitian quantum mechanics involve
the study of scattering by complex potentials and
quantum transport~\cite{suu54,ck58,lay63,bender07,varga,berg,miro,varga2,muga,thila,wahlstrand2014wave},
description of metastable states~\cite{nimrod2,seba,spyros,fesh,fesh2,sudarshan},
optical waveguides~\cite{optics,optics2,schomerus}, 
multi-photon ionization~\cite{selsto,baker2,chu}, and nano-photonic and plasmonic
waveguides~\cite{alaeian2014no}.
The theoretical investigations are also undergoing rapid developments:
non-Hermitian quantum mechanics has been investigated within a
relativistic framework~\cite{jones-smith} and it has been adopted
by various researchers as a means to describe open quantum 
systems~\cite{kor64,wong67,heg93,bas93,ang95,rotter2,bellomo,banerjee,reiter,bg12,askz-lindblad}.
Moreover, it seems that a few theoretical studies have been dedicated to
the statistical mechanics and dynamics of systems with non-Hermitian
Hamiltonians~\cite{baker,dattoli,hu,eva-dimer,eva-roman,jones,sergi-matrix,askz,askz-cor,Kawamoto,lian2014mathcal,karakaya2014non}.

In the present work, the interest is focused on the development of
a formalism to embed consistently the quantum dynamics of systems with a non-conserved 
probability in a classical environment, which is explicitly taken
into account (\emph{i.e.}, it is not averaged-over) in the dynamics
and which , in turn, acts as a source of disorder.
Types of noise beyond those
arising from Gaussian white noise~\cite{bg12} can then be treated.
From a more general perspective, one goal of this work is to develop
a numerical formalism 
(which is complementary to that based on master equations~\cite{bp})
for studying the dissipative dynamics of, for example,
quantum plasmonic metamaterials~\cite{tame,tame2} or
processes of interest in quantum thermodynamics~\cite{mahler}.
In order to obtain such a formalism, a composite system
with heavy and light degrees of freedom is considered.
First a classical limit over the heavy degrees of freedom is performed
using the partial Wigner representation~\cite{osborn,balescu,boucher,stock,martens,mqc,sk-rate,theor,sk-cor}.
In such a way, a general quantum-classical approximation of
non-Hermitian quantum mechanics is obtained.
Finally, the limiting case in which the non-Hermitian
part of the evolution does not affect the classical-like degrees
of freedom (represented in Wigner phase space~\cite{wigner,wigner2}) is considered.

This paper is structured as follows.
In Sec.~\ref{sec:qcnndm-dyna} the quantum non-Hermitian equation of motion
for the density matrix is taken as a starting point and the classical limit
over the heavy mass coordinates is performed by using the partial Wigner
transform and the linear expansion in the square root of the ratio
between light and heavy masses.
From this general case, one can easily derive the equation of motion
valid for a decay operator
depending only on the quantum degrees of freedom of the subsystem.
In Sec.~\ref{sec:adbas_rep} 
the Hermitian part of the total Hamiltonian of the system
is considered; an adiabatic Hamiltonian is extracted from this
and its eigenstates are used to represent the quantum-classical
non-Hermitian equation.
In Sec.~\ref{sec:sh_alg} piecewise-deterministic algorithms
(using the adiabatic basis) 
are presented explicitly in the case when the decay operator depends
only on the quantum coordinates of the subsystem.
The numerical approach is illustrated in Sec.~\ref{sec:twospins}
by studying (in the case of a weak coupling to the environment)
the dynamics of a chain of two spins, coupled separately
to an independent harmonic oscillator.
Two different decay operators are explicitly treated.
The evolution of the trace of the reduced density matrix 
of the spin chain and of relevant matrix elements is monitored.
The results show that non-Hermitian quantum-classical dynamics (and the
numerical algorithms developed in this work) can model
the loss of probability and the damping expected in
open quantum system.
Conclusions are finally given in Sec.~\ref{sec:conclusions}.

\section{Non-Hermitian quantum mechanics in a classical bath}
\label{sec:qcnndm-dyna}

Consider a composite quantum system 
with quantum coordinates
$(\hat{r},\hat{p},\hat{R},\hat{P})=(\hat{x},\hat{X})$.
A multidimensional notation will be adopted in the following so
that, for example, $\hat{R}$ stands for $(\hat{R}_1,\hat{R}_2,...,\hat{R}_N$),
where $N$ is the total number of degrees of freedom in the configurational space
of the subsystem represented by the operators $\hat{X}$.
It is also assumed that the dynamics of the composite system 
is defined by the non-Hermitian Hamiltonian
\begin{equation}
\hat{\cal H}=\hat{H}-i\hat{\Gamma}\;,
\label{eq:H-iG}
\end{equation}
where $\hat{H}=(\hat{\cal H}+\hat{\cal H}^\dag)/2$ and
$\hat{\Gamma}=i(\hat{\cal H}-\hat{\cal H}^\dag)/2$ are Hermitian operators.
In particular, in order to develop the formalism, one can assume that
\begin{equation}
\hat{H}=\frac{\hat{P}^2}{2M}+\frac{\hat{p}^2}{2m}+\hat{V}(\hat{r},\hat{R})
\;. \label{eq:quant_H}
\end{equation}
while $\hat{\Gamma}$ is left unspecified.
In order to have an idea of the type of physical systems to which
the formalism arising from Eq.~(\ref{eq:H-iG}) can be applied,
one can consider, for example, a
quantum resonance of the following type: a double well with a decay
operator providing the metastability of the bound state.
The non-unitary dynamics would allow the particle to escape
from the bound state while eventual quantum tunneling 
would allow it to go
from one well to another. Such internal transitions
between the wells can be enhanced by the thermal disorder
provided by a classical environment coupled to the well.

In the following, it will also be assumed that $M$, the mass associated
to the sub-system with coordinates $\hat{X}$, is much bigger
than $m$, the mass associated to the subsystem with coordinates
$\hat{x}$, \emph{i.e.}, $M>>m$.
This leads to the definition of the small parameter $\mu=(m/M)^{1/2}$.
The non-normalized density matrix $\hat{\Omega}(t)$
of the composite system with non-Hermitian
Hamiltonian $\cal H$ obeys the equation of motion~\cite{eva-roman}
\begin{equation}
\frac{\partial}{\partial t}\hat{\Omega}(t)
=-\frac{i}{\hbar}\left[\hat{H},\hat{\Omega}(t)\right]_-
-\frac{1}{\hbar}\left[\hat{\Gamma},\hat{\Omega}(t)\right]_+
\;, \label{eq:ddt_Omega}
\end{equation}
where $[...,...]_-$ and $[...,...]_+$
are the commutator and anticommutator, respectively.

In order to obtain the quantum-classical limit of Eq.~(\ref{eq:ddt_Omega})
a rigorous procedure, which is based on the partial Wigner representation
of the dynamics and the linear expansion in $\mu$, can be followed.
Such a procedure was used for Hermitian Hamiltonians in Ref.~\cite{mqc}.
Accordingly, one can introduce the partial Wigner transform of $\hat{\Omega}$
over only the coordinates of the heavy degrees of freedom:
\begin{equation}
\hat{\Omega}_{\rm W}(X,t)
=\frac{1}{(2\pi\hbar)^N}\int dZ e^{i P\cdot Z/\hbar}
\langle R - Z/2|\hat{\Omega}(t)|R+Z/2\rangle\;.
\end{equation}
As a results $\hat{\Omega}_{\rm W}(X,t)$ is an operator in terms of
the quantum $\hat{x}$ variables and a function in terms of the $X$ variables
(which are still quantum - they have only been represented in the
Wigner quantum phase space).
Analogously, an arbitrary quantum operator $\hat{\chi}$ of the composite system
is partially transformed in Wigner phase space as
\begin{equation}
\hat{\chi}_{\rm W}(X)
=\int dZ e^{i P\cdot Z/\hbar} \langle R - Z/2|\hat{\chi}|R+Z/2\rangle\;.
\end{equation}
Moreover, the partial Wigner transform of a product of arbitrary operators
$\hat{\chi}$ and $\hat{\xi}$ is given by
\begin{equation}
\left(\hat{\chi}\hat{\xi}\right)_{\rm W}(X)
\equiv \hat{\chi}_{\rm W}(X) e^{\frac{i\hbar}{2}
\overleftarrow{\mbox{\boldmath$\nabla$}}
\mbox{\boldmath$\cal B$}
\overrightarrow{ \mbox{\boldmath$\nabla$} }
}
\hat{\xi}_{\rm W}(X)
\;,
\end{equation}
where $\mbox{\boldmath$\nabla$}=((\partial/\partial R),(\partial/\partial P))$
is the phase space gradient operator and
\begin{equation}
\mbox{\boldmath$\cal B$}=\left[\begin{array}{cc} 0 & 1 \\ -1 & 0\end{array}\right]
\end{equation}
is the symplectic matrix, so that
$
\overleftarrow{\mbox{\boldmath$\nabla$}}
\mbox{\boldmath$\cal B$}
\overrightarrow{ \mbox{\boldmath$\nabla$} }
$ is basically the Poisson bracket operator, which will also be denoted
with the symbol $\{...,...\}$.

Upon taking the partial Wigner transform of Eq.~(\ref{eq:ddt_Omega}),
one obtains
\begin{eqnarray}
\frac{\partial}{\partial t}\hat{\Omega}_{\rm W}(X,t)
=
&-&\frac{i}{\hbar}
\left(\hat{H}_{\rm W}(X)
e^{\frac{i\hbar}{2} \overleftarrow{\mbox{\boldmath$\nabla$}}
\mbox{\boldmath$\cal B$}
\overrightarrow{ \mbox{\boldmath$\nabla$} } }
\hat{\Omega}_{\rm W}(X,t)
-
\hat{\Omega}_{\rm W}(X,t)
e^{\frac{i\hbar}{2} \overleftarrow{\mbox{\boldmath$\nabla$}}
\mbox{\boldmath$\cal B$}
\overrightarrow{ \mbox{\boldmath$\nabla$} } }
\hat{H}_{\rm W}(X)
\right)
\nonumber\\
&-&\frac{1}{\hbar}
\left(\hat{\Gamma}_{\rm W}(X)
e^{\frac{i\hbar}{2} \overleftarrow{\mbox{\boldmath$\nabla$}}
\mbox{\boldmath$\cal B$}
\overrightarrow{ \mbox{\boldmath$\nabla$} } }
\hat{\Omega}_{\rm W}(X,t)
+
\hat{\Omega}_{\rm W}(X,t)
e^{\frac{i\hbar}{2} \overleftarrow{\mbox{\boldmath$\nabla$}}
\mbox{\boldmath$\cal B$}
\overrightarrow{ \mbox{\boldmath$\nabla$} } }
\hat{\Gamma}_{\rm W}(X)
\right)
\;. \nonumber\\
\label{eq:ddt_Omega_W}
\end{eqnarray}
Equation~(\ref{eq:ddt_Omega_W}) is still fully quantum in nature.
Now, in order to take the quantum classical limit one can follow
the procedure of Ref.~\cite{mqc}, which was inspired by
the theory of Brownian motion given in~\cite{bm},
and introduce scaled coordinates.
Arbitrary units can be defined in the following way: 
one can introduce $\epsilon_0$ as unit of energy,
$t_0=\hbar/\epsilon_0$ as unit of time, and 
$\lambda_0=\hbar/(m\epsilon_0)^{(1/2)}$ as unit of length.
One can also define $p_0=(m\epsilon_0)^{1/2}$ and 
$P_0=(M\epsilon_0)^{1/2}$ as the unit of the light and heavy momenta,
respectively. As in~\cite{mqc}, adimensional coordinates,
making the momenta of the same order of magnitude, can be introduced:
\begin{equation}
\begin{array}{ccccccc}
\hat{r}^{\prime} & = &\frac{\hat{r}}{\lambda_0}\;, &\phantom{lala} &
R^{\prime}&=&\frac{R}{\lambda_0}\;,
\\
\hat{p}^{\prime}&=&\frac{\hat{p}}{p_0}\;, &\phantom{lala} &
P^{\prime}&=&\frac{P}{P_0}\;.
\end{array}
\label{eq:scal_coord}
\end{equation}
Accordingly, one has
$\hat{H}_{\rm W}(\hat{x},X)=\epsilon_0\hat{H}_{\rm W}^{\prime}(\hat{x},X)$
and
$\hat{\Gamma}_{\rm W}(\hat{x},X)=\epsilon_0\hat{\Gamma}_{\rm W}^{\prime}(\hat{x},X)$.
In the scaled coordinates given in Eq.~(\ref{eq:scal_coord}),
Eq.~(\ref{eq:ddt_Omega_W}) becomes
\begin{eqnarray}
\frac{\partial}{\partial t'}\hat{\Omega}_{\rm W}^{\prime}(X',t')
=
&-&
i\left(\hat{H}_{\rm W}^{\prime}(X')
e^{\frac{i\mu}{2} \overleftarrow{\mbox{\boldmath$\nabla$}}^{\prime}
\mbox{\boldmath$\cal B$}
\overrightarrow{ \mbox{\boldmath$\nabla$} }^{\prime} }
\hat{\Omega}_{\rm W}^{\prime}(X',t')
\right. \nonumber\\
&-&
\left.
\hat{\Omega}_{\rm W}^{\prime}(X',t')
e^{\frac{i\mu}{2} \overleftarrow{\mbox{\boldmath$\nabla$}}^{\prime}
\mbox{\boldmath$\cal B$}
\overrightarrow{ \mbox{\boldmath$\nabla$} }^{\prime} }
\hat{H}_{\rm W}^{\prime}(X')
\right)
\nonumber\\
&-&
\left(\hat{\Gamma}_{\rm W}^{\prime}(X')
e^{\frac{i\mu}{2} \overleftarrow{\mbox{\boldmath$\nabla$}}^{\prime}
\mbox{\boldmath$\cal B$}
\overrightarrow{ \mbox{\boldmath$\nabla$} }^{\prime} }
\hat{\Omega}_{\rm W}^{\prime}(X',t.)
\right. \nonumber\\
&+&
\left.
\hat{\Omega}_{\rm W}^{\prime}(X',t')
e^{\frac{i\mu}{2} \overleftarrow{\mbox{\boldmath$\nabla$}}^{\prime}
\mbox{\boldmath$\cal B$}
\overrightarrow{ \mbox{\boldmath$\nabla$} }^{\prime} }
\hat{\Gamma}_{\rm W}^{\prime}(X')
\right)
\;, 
\label{eq:ddt_Omega_W_scal}
\end{eqnarray}
where $t'=t/t_0$.
Now one can take advantage of the smallness of $\mu$ and
expand the exponential operators retaining only the linear order terms.
One obtains
\begin{eqnarray}
\frac{\partial}{\partial t'}\hat{\Omega}_{\rm W}^{\prime}(X',t')
=
&-&i\left[\hat{H}_{\rm W}^{\prime}(X'),\hat{\Omega}_{\rm W}^{\prime}(X',t')\right]_-
-\left[\hat{\Gamma}_{\rm W}^{\prime}(X'),\hat{\Omega}_{\rm W}^{\prime}(X',t')\right]_+
\nonumber\\
&+&\sum_{jk}\left[\frac{\mu}{2}{\cal B}_{jk}
\left(\nabla_j^{\prime}\hat{H}_{\rm W}^{\prime}(X')\right)
\nabla_k^{\prime}\hat{\Omega}_{\rm W}^{\prime}(X',t')
\right.
\nonumber\\
&-&\frac{\mu}{2}{\cal B}_{jk}
\left(\nabla_j^{\prime}\hat{\Omega}_{\rm W}^{\prime}(X',t')\right)
\nabla_k^{\prime}\hat{H}_{\rm W}^{\prime}(X')
\nonumber\\
&-&\frac{i\mu}{2}{\cal B}_{jk}
\left(\nabla_j^{\prime}\hat{\Gamma}_{\rm W}^{\prime}(X')\right)
\nabla_k^{\prime}\hat{\Omega}_{\rm W}^{\prime}(X',t')
\nonumber\\
&-&\left.\frac{i\mu}{2}{\cal B}_{jk}
\left(\nabla_j^{\prime}\hat{\Omega}_{\rm W}^{\prime}(X',t')\right)
\nabla_k^{\prime}\hat{\Gamma}_{\rm W}^{\prime}(X')
\right]
\;, \label{eq:qc_ddt_Omega_W_scal}
\end{eqnarray}
where the sum over the indices $j,k$ runs over all phase space dimensions.
Equation~(\ref{eq:qc_ddt_Omega_W_scal}) gives the 
quantum-classical approximation
to Eqs.~(\ref{eq:ddt_Omega})  and (\ref{eq:ddt_Omega_W}) in adimensional coordinates.
Transforming back to fully dimensional variables, one finally obtains
\begin{eqnarray}
\frac{\partial}{\partial t}\hat{\Omega}_{\rm W}(X,t)
=
&-&\frac{i}{\hbar}\left[\hat{H}_{\rm W}(X),\hat{\Omega}_{\rm W}(X,t)\right]_-
-\frac{1}{\hbar}\left[\hat{\Gamma}_{\rm W}(X),\hat{\Omega}_{\rm W}(X,t)\right]_+
\nonumber\\
&+&\frac{1}{2}\left(
\left\{\hat{H}_{\rm W}(X),\hat{\Omega}_{\rm W}(X,t)\right\}
-\left\{\hat{\Omega}_{\rm W}(X,t),\hat{H}_{\rm W}(X)\right\}
\right)
\nonumber\\
&-&\frac{i}{2}
\left(
\left\{\hat{\Gamma}_{\rm W}(X),\hat{\Omega}_{\rm W}(X,t)\right\}
+\left\{\hat{\Omega}_{\rm W}(X,t),\hat{\Gamma}_{\rm W}(X)\right\}
\right)
\;.  \label{eq:qc_ddt_Omega_W}
\end{eqnarray}
Equation~(\ref{eq:qc_ddt_Omega_W}) provides
the rigorous quantum-classical approximation
for the non-Hermitian dynamics of composite systems. It is a valid approximation
when the degrees of freedom of the system have two different
De Broglie wavelengths, one short and one long.
It is worth noting that, formally, Eq.~(\ref{eq:qc_ddt_Omega_W})
could have been obtained directly from 
Eq.~(\ref{eq:ddt_Omega_W}) by taking an expansion to linear order in the
limit $\hbar\to 0$, without going through the transformation
to scaled coordinates given in Eq.~(\ref{eq:scal_coord}).
However, the $\mu$ expansion seems much more rigorous since $\hbar\to 0$
would imply a cancelation of quantum effects also on the $\hat{x}$
coordinates. Nevertheless, the practical agreement of the two limiting
procedures indicates that the linear form in Eq.~(\ref{eq:qc_ddt_Omega_W})
can perhaps simply assumed as an ansatz for quantum-classical dynamics,
as suggested in Ref.~\cite{b4}.
It is also very important to remark that Eq.~(\ref{eq:qc_ddt_Omega_W})
has a form that does not depend on any particular basis.
Moreover, Eq.~(\ref{eq:qc_ddt_Omega_W}) is exact when the decay operator
$\hat{\Gamma}_{\rm W}(X)$ is linear in $X$, $\hat{H}_{\rm W}(X)$
is at most quadratic in the $X$ coordinates and is linearly coupled
through the $X$s with the quantum subsystem.


While Eq.~(\ref{eq:qc_ddt_Omega_W}) defines non-Hermitian quantum dynamics
in a classical bath in the case of a general decay operator $\hat{\Gamma}_{\rm W}(X)$,
there is one interesting limiting situation that can be considered.
It concerns the case in which the decay operator does not depend on 
the bath coordinates. When this happens the partial Wigner transform
leaves $\hat{\Gamma}$ invariant, so that Eq.~(\ref{eq:qc_ddt_Omega_W}) reduces to
\begin{eqnarray}
\frac{\partial}{\partial t}\hat{\Omega}_{\rm W}(X,t)
=
&-&\frac{i}{\hbar}\left[\hat{H}_{\rm W}(X),\hat{\Omega}_{\rm W}(X,t)\right]_-
-\frac{1}{\hbar}\left[\hat{\Gamma},\hat{\Omega}_{\rm W}(X,t)\right]_+
\nonumber\\
&+&\frac{1}{2}\left(
\left\{\hat{H}_{\rm W}(X),\hat{\Omega}_{\rm W}(X,t)\right\}
-\left\{\hat{\Omega}_{\rm W}(X,t),\hat{H}_{\rm W}(X)\right\}
\right)
\;. \label{eq:qc_ddt_Omega_W_quant_Gamma}
\end{eqnarray}
Equation~(\ref{eq:qc_ddt_Omega_W_quant_Gamma}) shows that,
in this case, the effects on the subsystem dynamics
arise from the anticommutator of $\hat{\Gamma}$ and $\hat{\Omega}_{\rm W}(X)$
alone. Equation~(\ref{eq:qc_ddt_Omega_W_quant_Gamma}) might describe a situation
in which two types of effects are present: the non-Hermitian dynamics, with its
probability leakage or pumping, of a quantum subsystem
embedded in a bath of classical degrees of freedom, whose influence is expressed
through the Poisson bracket terms on the right hand side.
If also the density matrix does not depend
on the bath coordinates $X$,
$\hat{\Omega}_{\rm W}(X,t)\to\hat{\Omega}(t)$, one obtains
the purely quantum case, given by Eq.~(\ref{eq:ddt_Omega}).
Such a result constitutes a self-consistency check for
the formalism.

\section{Representation in the adiabatic basis}
\label{sec:adbas_rep}

The partial Wigner transform of the
Hamiltonian in Eq.~(\ref{eq:quant_H}) can be rewritten as
\begin{equation}
\hat{H}_{\rm W}(X)=\frac{P^2}{2M}+\hat{h}_{\rm W}(R)\;.
\end{equation}
The adiabatic basis of $\hat{H}_{\rm W}(X)$ is defined by the eigenvalue problem
\begin{equation}
\hat{h}_{\rm W}|\alpha;R\rangle=E_{\alpha}(R)|\alpha;R\rangle\;.
\end{equation}
Such a basis can be used to represent Eq.~(\ref{eq:qc_ddt_Omega_W}).
Upon defining the quantities
\begin{eqnarray}
\Omega_{\rm W}^{\alpha\alpha'}(X,t)
&=&\langle\alpha;R|\hat{\Omega}_{\rm W}(X,t)
|\alpha';R\rangle
\;, \\
\omega_{\alpha\alpha'}&=&\frac{E_{\alpha}(R)-E_{\alpha'}(R)}{\hbar}
\label{eq:bohr_freq}
\;, \\
F_{\rm W}^\alpha&=&-\frac{\partial E_{\alpha}(R)}{\partial R}
\;, \\
iL_{\alpha\alpha'}&=&\frac{P}{M}\cdot\frac{\partial}{\partial R}
+\frac{1}{2}(F_{\rm W}^\alpha+F_{\rm W}^{\alpha'})\cdot
\frac{\partial}{\partial P}
\label{eq:L_def}
\;,\\
d_{\alpha\alpha'}&=&\langle\alpha;R|\frac{\partial}{\partial R}|\alpha';R\rangle
\;,\\
S_{\alpha\beta}&=&\left(\frac{P}{M}\cdot d_{\alpha\beta}\right)^{-1}
\hbar\omega_{\alpha\beta}d_{\alpha\beta}
\;,\\
{\cal T}_{\alpha\alpha',\beta\beta'}
&=&
\delta_{\alpha'\beta'}\frac{P}{M}\cdot d_{\alpha\beta}
\left(1+\frac{1}{2}S_{\alpha\beta}\cdot\frac{\partial}{\partial P}\right)
\nonumber\\
&+&
\delta_{\alpha\beta}\frac{P}{M}\cdot d_{\alpha'\beta'}^*
\left(1+\frac{1}{2}S_{\alpha'\beta'}^*\cdot\frac{\partial}{\partial P}\right)
\label{eq:cal_Tau_def}
\;,\\
i{\cal L}_{\alpha\alpha',\beta\beta'}^0
&=&\left(i\omega_{\alpha\alpha'}+iL_{\alpha\alpha'}\right)
\delta_{\alpha\beta}\delta_{\alpha'\beta'} \label{eq:cal_L0_def}
\;,\\
i{\cal L}_{\alpha\alpha',\beta\beta'}&=&i{\cal L}_{\alpha\alpha',\beta\beta'}^{(0)}
+{\cal T}_{\alpha\alpha',\beta\beta'}\;, \label{eq:cal_L_def}
\end{eqnarray}
and using the results of section III in Ref.~\cite{mqc},
providing the representation of
the first, third and fourth terms
in the right hand side of Eq.~(\ref{eq:qc_ddt_Omega_W}),
one obtains for Eq.~(\ref{eq:qc_ddt_Omega_W}):
\begin{eqnarray}
\frac{\partial}{\partial t}\Omega_{\rm W}^{\alpha\alpha'}
=
&-&\sum_{\beta\beta'}i{\cal L}_{\alpha\alpha',\beta\beta'}
\Omega_{\rm W}^{\beta\beta'}
-\frac{1}{\hbar}\langle\alpha;R|\left[\hat{\Gamma}_{\rm W},\hat{\Omega}_{\rm W}(X,t)
\right]_+|\alpha';R\rangle
\nonumber\\
&-&\frac{i}{2}
\langle\alpha;R|\left(\left\{\hat{\Gamma}_{\rm W},\hat{\Omega}_{\rm W}(X,t)\right\}
+\left\{\hat{\Omega}_{\rm W}(X,t),\hat{\Gamma}_{\rm W}\right\}\right)
|\alpha';R\rangle
\;.
\label{eq:Omega_W_ad_rep_1st}
\end{eqnarray}
To proceed one can take advantage of the identities
\begin{eqnarray}
\langle\alpha;R|\frac{\partial\hat{\Gamma}_{\rm W}}{\partial R}
|\gamma;R\rangle
&=&
\frac{\partial\Gamma_{\rm W}^{\alpha\gamma}}{\partial R}
-\sum_{\sigma} d_{\sigma\alpha}^* \Gamma_{\rm W}^{\sigma\gamma}
-\sum_{\sigma} \Gamma_{\rm W}^{\alpha\sigma} d_{\sigma\gamma}
\;, \label{eq:iden1}\\
\langle\gamma;R|
\frac{\partial\hat{\Omega}_{\rm W}(X,t)}{\partial R}
|\alpha';R\rangle
&=&
\frac{\partial}{\partial R}\Omega_{\rm W}^{\gamma\alpha'}
-\sum_{\sigma}d_{\sigma\gamma}^*\Omega_{\rm W}^{\sigma\alpha'}
-\sum_{\sigma}d_{\sigma\alpha'}\Omega_{\rm W}^{\gamma\sigma}
\;.\label{eq:iden2}
\end{eqnarray}
The derivation is lengthy but straightforward.
Its details are provided in App.~\ref{app:derivation}.
The representation of Eq.~(\ref{eq:qc_ddt_Omega_W_quant_Gamma})
into the adiabatic basis is
\begin{eqnarray}
\frac{\partial}{\partial t}\Omega_{\rm W}^{\alpha\alpha'}
=
&-&\sum_{\beta\beta'}i{\cal L}_{\alpha\alpha',\beta\beta'}
\Omega_{\rm W}^{\beta\beta'}
-\frac{1}{\hbar}\sum_{\beta\beta'}
\left( \Gamma_{\rm W}^{\alpha\beta}\delta_{\alpha'\beta'}
+ \Gamma_{\rm W}^{\beta'\alpha'}\delta_{\alpha\beta} \right)
\Omega_{\rm W}^{\beta\beta'}(X,t)
\nonumber\\
&-&\frac{i}{2}
\sum_{\beta\beta'}
\left[ \left(
\frac{\partial\Gamma_{\rm W}^{\alpha\beta}}{\partial R}
\delta_{\alpha'\beta'} 
- \frac{\partial\Gamma_{\rm W}^{\beta'\alpha'}}{\partial R}
\delta_{\alpha\beta}\right)
\frac{\partial}{\partial P}
\right.\nonumber\\
&+&
\left.
\left( \frac{\partial\hat{\Gamma}_{\rm W}^{\beta'\alpha'}}{\partial P}
\delta_{\alpha\beta}
- \frac{\partial\Gamma_{\rm W}^{\alpha\beta}}{\partial P}
\delta_{\alpha'\beta'}\right)\frac{\partial}{\partial R}
\right.
\nonumber\\
&-&
\sum_{\sigma}\left(
d_{\sigma\alpha}^* \Gamma_{\rm W}^{\sigma\beta}\delta_{\alpha'\beta'}
+
\Gamma_{\rm W}^{\alpha\sigma} d_{\sigma\beta} \delta_{\alpha'\beta'}
\right.\nonumber\\
&+&
\left.
d_{\sigma\beta'}^*\Gamma_{\rm W}^{\sigma\alpha'}
\delta_{\alpha\beta}
+
\Gamma_{\rm W}^{\beta'\sigma}d_{\sigma\alpha'}
\delta_{\alpha\beta}
\right)
\frac{\partial}{\partial P}
\nonumber\\
&+&
\left.
\sum_{\sigma}\left(
\frac{\partial\Gamma_{\rm W}^{\alpha\sigma}}{\partial P}
d_{\beta\sigma}^*\delta_{\alpha'\beta'}
-d_{\beta'\sigma}
\frac{\partial\hat{\Gamma}_{\rm W}^{\sigma\alpha'}}{\partial P}
\delta_{\alpha\beta}
\right)
\right.\nonumber\\
&+&
\left.
\frac{\partial\Gamma_{\rm W}^{\alpha\beta}}{\partial P}
d_{\beta'\alpha'}
-
d_{\beta\alpha}^*
\frac{\partial\hat{\Gamma}_{\rm W}^{\beta'\alpha'}}{\partial P}
\right]
\hat{\Omega}_{\rm W}^{\beta\beta'}
\;.\label{eq:Omega_W_ad_rep_4th}
\end{eqnarray}
Equation~(\ref{eq:Omega_W_ad_rep_4th}) provides the desired result:
it is the representation of the quantum-classical non-Hermitian
dynamics, given by Eq.~(\ref{eq:qc_ddt_Omega_W}), in the
adiabatic basis of $\hat{H}_{\rm W}(X)$.
As it stands, it looks quite formidable and not amenable of
being numerically integrated in a simple way.

When the decay operator depends only on the quantum coordinates
of the subsystem,
one has to consider the representation of
Eq.~(\ref{eq:qc_ddt_Omega_W_quant_Gamma}) in the adiabatic basis.
It can be verified that Eq.~(\ref{eq:Omega_W_ad_rep_4th}) reduces to
\begin{eqnarray}
\frac{\partial}{\partial t}\Omega_{\rm W}^{\alpha\alpha'}(X,t)
=
&-&\sum_{\beta\beta'}i{\cal L}_{\alpha\alpha',\beta\beta'}
\Omega_{\rm W}^{\beta\beta'}(X,t)
\nonumber\\
&-&\frac{1}{\hbar}\sum_{\beta\beta'}
\left( \Gamma^{\alpha\beta}\delta_{\alpha'\beta'}
+ \Gamma^{\beta'\alpha'}\delta_{\alpha\beta} \right)
\Omega_{\rm W}^{\beta\beta'}(X,t) 
\;.  \label{eq:Omega_W_ad_rep_quant_Gamma}
\end{eqnarray}
Equation~(\ref{eq:Omega_W_ad_rep_quant_Gamma})
is much simpler than the general Eq.~(\ref{eq:Omega_W_ad_rep_4th});
it can be used to model dissipative effects on
quantum subsystems arising both from sources and sinks of 
probability and disorder because of the coupling
to a classical environment.
Equation~(\ref{eq:Omega_W_ad_rep_quant_Gamma})
can be integrated by means of piecewise-deterministic algorithms.
These will be sketched in Sec.~\ref{sec:sh_alg}.


\section{Piecewise-deterministic algorithms}
\label{sec:sh_alg}

Equation~(\ref{eq:Omega_W_ad_rep_4th}), which is 
the adiabatic representation of the abstract Eq.~(\ref{eq:qc_ddt_Omega_W}),
can be numerically integrated by means of piecewise-deterministic algorithms. 
However, given the complexity of Eq.~(\ref{eq:Omega_W_ad_rep_4th}),
only the limiting case
given by Eq.~(\ref{eq:Omega_W_ad_rep_quant_Gamma})
will be explicitly considered here.

Consider first Eq.~(\ref{eq:Omega_W_ad_rep_quant_Gamma}).
In such a case, it is convenient to decompose the representation
of the decay operator in the adiabatic basis in terms of a
diagonal, $\Gamma_{\rm d}^{\alpha\alpha}$,
and an off-diagonal part, $\Gamma_{\rm o}^{\alpha\beta}$:
\begin{equation}
\Gamma^{\alpha\beta}=\Gamma_{\rm d}^{\alpha\alpha}\delta_{\alpha\beta}
+\Gamma_{\rm o}^{\alpha\beta} \;.
\end{equation}
Equation~(\ref{eq:Omega_W_ad_rep_quant_Gamma}) becomes
\begin{eqnarray}
\frac{\partial}{\partial t}\Omega_{\rm W}^{\alpha\alpha'}
&=&
-\sum_{\beta\beta'}i{\cal L}_{\alpha\alpha',\beta\beta'}
\Omega_{\rm W}^{\beta\beta'}
-\frac{1}{\hbar}\sum_{\beta\beta'}
\left( \Gamma^{\alpha\alpha}_{\rm d}
+ \Gamma_{\rm d}^{\alpha'\alpha'} \right)\delta_{\alpha'\beta'}\delta_{\alpha\beta}
\Omega_{\rm W}^{\beta\beta'}(X,t)
\nonumber\\
&&
-\frac{1}{\hbar}\sum_{\beta\beta'}
\left( \Gamma_{\rm o}^{\alpha\beta}\delta_{\alpha'\beta'}
+ \Gamma_{\rm o}^{\beta'\alpha'}\delta_{\alpha\beta} \right)
\Omega_{\rm W}^{\beta\beta'}(X,t)
\;. \label{eq:Omega_W_ad_rep_quant_Gamma_2nd}
\end{eqnarray}
At this point, it is useful to define
\begin{eqnarray}
\gamma_{\alpha\alpha'}
&=&\frac{\Gamma^{\alpha\alpha}_{\rm d}+\Gamma_{\rm d}^{\alpha'\alpha'}}{\hbar}
\label{eq:gamma-alphaalphap}\;, \\
{\cal T}_{\alpha\alpha',\beta\beta'}^{\Gamma}
&=&\frac{1}{\hbar}\left(\Gamma_{\rm o}§^{\alpha\beta}\delta_{\alpha'\beta'}
+ \Gamma_{\rm o}^{\beta'\alpha'}\delta_{\alpha\beta}\right) \;.
\end{eqnarray}
Equation~(\ref{eq:Omega_W_ad_rep_quant_Gamma_2nd}) becomes
\begin{eqnarray}
\frac{\partial}{\partial t}\Omega_{\rm W}^{\alpha\alpha'}
&=&
-\sum_{\beta\beta'}\left(
i{\cal L}_{\alpha\alpha',\beta\beta'}^{(\gamma)}
+ {\cal T}_{\alpha\alpha',\beta\beta'}+ 
{\cal T}_{\alpha\alpha',\beta\beta'}^{\Gamma}
\right)
\Omega_{\rm W}^{\beta\beta'}(X,t)
\;,\label{eq:Omega_W_ad_rep_quant_Gamma_3rd}
\end{eqnarray}
where one has defined
\begin{equation}
i{\cal L}_{\alpha\alpha',\beta\beta'}^{(\gamma)}
=\left(i\omega_{\alpha\alpha'}
+ \gamma_{\alpha\alpha'}+ i L_{\alpha\alpha'} \right)\delta_{\alpha\beta}
\delta_{\alpha'\beta'}
\;.
\end{equation}
A piecewise-deterministic algorithm for the integration of
Eq.~(\ref{eq:Omega_W_ad_rep_quant_Gamma_3rd}) can be found by using the 
sequential short-time propagation (SSTP) scheme~\cite{theor,slice}.
A trajectory can be seen as the concatenation of small finite time steps $\Delta t$.
Accordingly, for a single step, the propagator associated with
Eq.~(\ref{eq:Omega_W_ad_rep_quant_Gamma_3rd}) can be written as
\begin{equation}
\left( e^{-i\Delta t\left({\cal L}^{(\gamma)}+{\cal T}+{\cal T}^{\Gamma}\right)}
\right)_{\alpha\alpha',\beta\beta'}
\approx
e^{-i\Delta t{\cal L}_{\alpha\alpha'}^{(\gamma)}}
\left(\delta_{\alpha\beta}\delta_{\alpha'\beta'}
-\Delta t {\cal T}_{\alpha\alpha',\beta'\beta'}
-\Delta t{\cal T}_{\alpha\alpha',\beta\beta'}^{\Gamma}\right)
\;. \label{eq:sstp_quant_Gamma}
\end{equation}
The propagator decomposition in Eq.~(\ref{eq:sstp_quant_Gamma})
can be used as the basis for a SSTP
algorithm for integrating Eq.~(\ref{eq:Omega_W_ad_rep_quant_Gamma_3rd}).
The actions of ${\cal T}_{\alpha\beta}\delta_{\alpha'\beta'}$
and ${\cal T}_{\alpha\beta}^{\Gamma}\delta_{\alpha'\beta'}$
must be sampled probabilistically using either basic~\cite{theor,slice}
or more advanced schemes~\cite{lt-sstpa,lt-sstpa2,lt-sstpa3}
for efficient convergence.
The momentum-jump approximation~\cite{theor,mj}
can be adopted in the expression of
${\cal T}_{\alpha\alpha',\beta\beta'}$.
It should be noted that the damping (or enhancing)
\emph{frequency} $\gamma_{\alpha\alpha'}$ is considered in the
action of $i{\cal L}_{\alpha\alpha'}^{(\gamma)}$.


\section{Non-Hermitian spin chain in harmonic baths}
\label{sec:twospins}

Consider a subsystem described by the Hermitian Hamiltonian
\begin{eqnarray}
\hat{H}_{\rm S}
&=&
-j_x\hat{\sigma}_x^{({\rm s}_1)}\hat{\sigma}_x^{({\rm s}_2)}
-j_y\hat{\sigma}_y^{({\rm s}_1)}\hat{\sigma}_y^{({\rm s}_2)}
-j_z\hat{\sigma}_z^{({\rm s}_1)}\hat{\sigma}_z^{({\rm s}_2)}
\;, \label{eq:H_S}
\end{eqnarray}
which represents a chain of two coupled quantum spins, ${\rm s}_k$,
$k=1,2$.
The constants $j_\ell$, with $\ell=x,y,z$,
determine the spin coupling strength.
The operators $\hat{\sigma}_\ell^{k_s}$ are the given
by the Pauli matrices for spin $k_s=1,2$.
The excited and ground state of the spins are denoted by $|e^{({\rm s}_k)}>$
and $|g^{({\rm s}_k)}>$ $(k=1,2)$, respectively.
As in Ref.~\cite{ilya}, the subsystem basis is 
defined by the following vectors: 
$|1>=|e^{({\rm s}_1)},e^{({\rm s}_2)}>$,
$|2>=|e^{({\rm s}_1)},g^{({\rm s}_2)}>$,
$|3>=|g^{({\rm s}_1)},e^{({\rm s}_2)}>$,
$|4>=|g^{({\rm s}_1)},g^{({\rm s}_2)}>$.
The bath is composed by two harmonic oscillators and has
the following partially Wigner-transformed Hamiltonian:
\begin{equation}
H_{\rm B, W}
=\sum_{k=1}^2\left(\frac{P_{({\rm s}_k)}^2}{2M}
+\frac{M\omega^2}{2}R_{({\rm s}_k)}^2\right)
\;. \label{eq:2spins_baths}
\end{equation}
Equation~(\ref{eq:2spins_baths}) provides the Wigner-transformed
Hamiltonian of two independent harmonic oscillators
with mass $M$ and frequency $\omega$.
Oscillator $1$ is coupled to spin $1$ while oscillator $2$
is coupled to spin $2$. The coupling Hamiltonian (in the
partial Wigner representation) is
\begin{equation}
\hat{H}_{\rm SB,W}=
-\sum_{k=1}^2cR_{({\rm s}_k)}\hat{\sigma}_z^{({\rm s}_k)}\;.
\label{eq:H_SB}
\end{equation}
The total partially Wigner-transformed Hermitian Hamiltonian
of the system is (of course) given by
$\hat{H}_{\rm W}(X)=\hat{H}_{\rm S}+\hat{H}_{\rm SB,W}+H_{\rm B, W}$.
Since the total bath is harmonic and the coupling with the spin chain
is bilinear, the linear approximation of the partially Wigner
represented dynamics is exact. This means that the classical-like
representation of the bath in Wigner space is, in fact,
fully quantum in nature.

In order to illustrate the numerical implementation of the formalism,
two decay operators are considered:
\begin{eqnarray}
\hat{\Gamma}^{(1)}&=&\gamma_1 \hat{I}\;,
\\
\hat{\Gamma}^{(2)}&=&\gamma_2
|e^{(s_1)},e^{(s_2)}><e^{(s_1)},e^{(s_2)}|\;.
\end{eqnarray} 
The symbol $\hat{I}$ denotes the identity operator in the Hilbert space 
of the spin chain while $\gamma_j$, $j=1,2$, are constants.
The operators are chosen so that the difference between
the effect of a uniform probability sink on all states of the
spin-chain, represented by operator $\hat{\Gamma}^{(1)}$,
and the depletion of just the one state (when both
spins are excited), represented by operator $\hat{\Gamma}^{(2)}$,
can be observed.
The dynamics of the density matrix is determined 
by substituting $\hat{\Gamma}^{(j)}$, $j=1,2$,
into Eq.~(\ref{eq:qc_ddt_Omega_W_quant_Gamma}).
The adiabatic basis representation of Eqs.~(\ref{eq:qc_ddt_Omega_W_quant_Gamma})
has been given in Sec.~\ref{sec:adbas_rep}.
Non-adiabatic corrections to the dynamics can be disregarded
upon assuming a weak coupling to the environment:
This means that the transition operators
${\cal T}_{\alpha\alpha',\beta\beta'}$
and
${\cal T}_{\alpha\alpha',\beta\beta'}^\Gamma$ 
in Eq.~(\ref{eq:Omega_W_ad_rep_quant_Gamma_3rd}) are neglected
in the calculations here discussed.
If $\Delta t$ is the numerical integration step,
a single-step  SSTP propagator
in the adiabatic approximation is written as
\begin{equation}
e^{-i\Delta t{\cal L}_{\alpha\alpha'}^{(\gamma)}}
=
e^{-i\int_0^{\Delta t}d\tau\omega_{\alpha\alpha'}(\tau)}
e^{-\frac{1}{\hbar}\int_0^{\Delta t} d\tau \gamma_{\alpha\alpha'}(\tau)}
e^{-i\Delta t L_{\alpha\alpha'}}
\;. \label{eq:sstp_class_Gamma_example}
\end{equation}
where $\gamma_{\alpha\alpha'}$ is defined in Eq.~(\ref{eq:gamma-alphaalphap}).
The right hand side of Eq.~(\ref{eq:sstp_class_Gamma_example})
can be derived by means of the Dyson identity,
as explained in Ref.~\cite{mqc}.
In the calculations reported either $\hat{\Gamma}^{(1)}$
or $\hat{\Gamma}^{(2)}$ 
have been used to obtain $\gamma_{\alpha\alpha'}$,
depending on the case.
\begin{figure}
\resizebox{7cm}{5cm}{
\includegraphics* {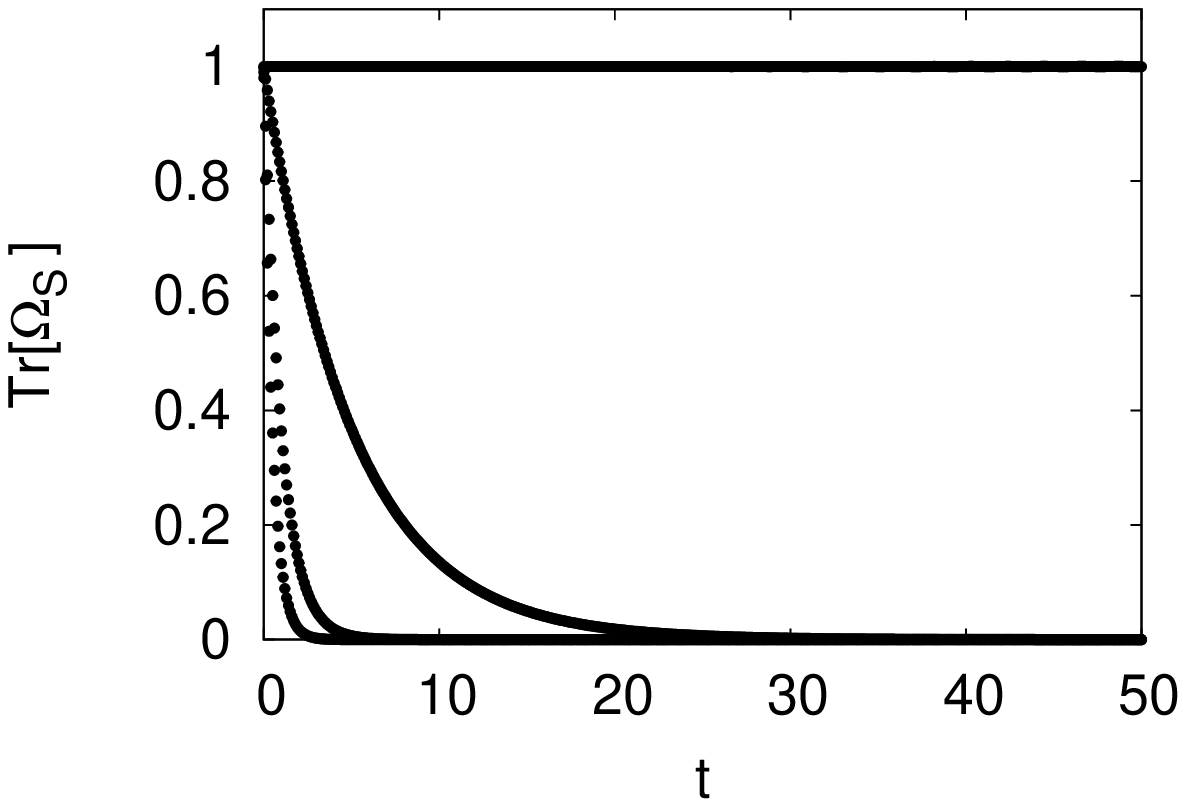}
}
\caption{
Adiabatic time evolution of the trace of the reduced density matrix,
${\rm Tr}[\Omega_{\rm S}]$,
for $\beta=0.1$, $j_x=j_y=-1$, $j_z=0.5$, $c=0.24$.
The numerical time step of integration is $\Delta t=0.01$.
Adimensional parameters are used.
Results for non-Hermitian dynamics with decay operator
$\hat{\Gamma}^{(1)}=\gamma_1\hat{1}$
and initial reduced density matrix
$\hat{\Omega}_{\rm S}(t_0)=|\Phi><\Phi|$, with 
$|\Phi>=|e^{(s_1)},g^{(s_2)}>$.
The upper curve  shows the results for $\gamma_1=0$ (Hermitian dynamics).
Then, from top to bottom, the curves for $\gamma_1=0.1, 0.5, 1$ are
displayed.
All curves are drawn with statistical error bars.
}
\label{fig:fig1}
\end{figure}

In order to perform the numerical study,
the density matrices of the subsystem $\hat{\Omega}_{\rm S}(t)$
and of the oscillators $\Omega_{\rm B,W}(X,t)$ are considered
uncorrelated at the inital time $t_{\rm i}$:
\begin{equation}
\hat{\Omega}_{\rm W}(X,t_{\rm i})
=
\hat{\Omega}_{\rm S}(t_{\rm i})\otimes \Omega_{\rm B,W}(X,t_{\rm i})\;,
\end{equation}
where
\begin{eqnarray}
\Omega_{\rm B,W}(X,t_{\rm i})
&=&
\prod_{k_s=1}^2\frac{\tanh(\beta\omega/2)}{\pi}
\exp\left[-2\frac{\tanh(\beta\omega/2)}{\omega}
H_{\rm B,W}(X)\right]\;,
\end{eqnarray}
with $\beta=1/k_{\rm B}T$ inverse thermodynamics temperature
($k_{\rm B}$ denotes the Boltzmann constant)
and $H_{\rm B,W}(X)$ is defined in Eq.~(\ref{eq:2spins_baths}).
The initial condition for the reduced density matrix of the spin chain
has been chosen as
$\hat{\Omega}_{\rm S}(t_{\rm i})=|\Phi><\Phi|$
with $|\Phi>=|e^{(s_1)},g^{(s_2)}\rangle$ when using $\hat{\Gamma}^{(1)}$ and as
$\hat{\Omega}_{\rm S}(t_{\rm i})=|\Psi><\Psi|$,
with
$|\Psi>=(1/\sqrt{2})\left(|e^{(s_1)},e^{(s_2)}>-|e^{(s_1)},g^{(s_2)}>\right)$
when using $\hat{\Gamma}^{(2)}$.
Upon choosing $\omega$ as a reference frequency, one can introduce a unit of
energy, $\hbar\omega$, a dimensionless time, $t\to \omega t$, and an
inverse thermodynamical temperature, $\beta\to\hbar\omega\beta$.
One can also introduce the adimensional coordinates $R\to(\hbar/M\omega)^{1/2}R$
and $P\to(\hbar\omega M)^{1/2}P$. Correspondingly, one can use the
following adimensional constants $j_\ell\to j_\ell/\hbar\omega$ ($\ell=x,y,z$),
$c\to c/(\hbar M\omega^2)$,
$\gamma_k\to\gamma_k/\hbar\omega$ ($k=1,2$).
The values adopted in the calculations have been $\beta=0.1$, 
$j_x=j_y=-1$, $j_z=0.5$, and $c=0.24$.
The parameters specifying the decay operators have been set,
in different calculations, to $\gamma_1=0.1,0.5,1$ and
$\gamma_2=0.001,0.01,0.1$. 
\begin{figure}
\resizebox{7cm}{5cm}{
\includegraphics* {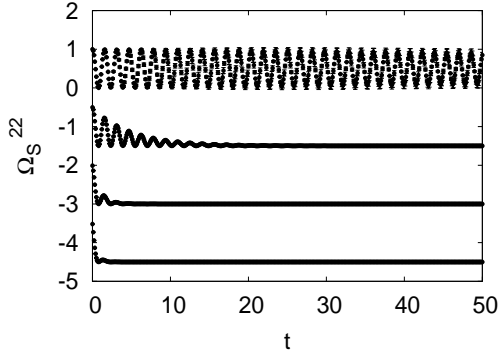}
}
\caption{Adiabatic time evolution of matrix element $\Omega_{\rm S}^{22}$,
in the subsystem basis, 
for $\beta=0.1$, $j_x=j_y=-1$, $j_z=0.5$, $c=0.24$.
The numerical time step of integration is $\Delta t=0.01$.
Adimensional parameters are used. 
Results for non-Hermitian dynamics with decay operator
$\Gamma_{\rm W}^{(1)}=\gamma_1\hat{1}$ and initial reduced density matrix
$\hat{\Omega}_{\rm S}(t_0)=|\Phi><\Phi|$, with 
$|\Phi>=|e^{(s_1)},g^{(s_2)}>$.
The upper curve displays the results for $\gamma_1=0$ (Hermitian Dynamics).
Then, from top to bottom, the curves for $\gamma_1=0.1, 0.5, 1$ are
displayed.
Starting from the top curve, a constant shift of 1.5 in the negative $y$
direction has been applied for visualization purposes.
All curves are drawn with statistical error bars.
}
\label{fig:fig2}
\end{figure}
Phase space averages with negligible statistical errors
have been calculated using $5\times 10^4$ points.

Figure~\ref{fig:fig1} displays the 
adiabatic time evolution of the trace of the reduced density matrix
of the spin chain, ${\rm Tr}[\Omega_{S}]$,
when the decay operator is $\hat{\Gamma}^{(1)}=\gamma_1\hat{1}$.
and initial reduced density matrix
$\hat{\Omega}_{\rm S}(0)=|\Phi><\Phi|$, with
$|\Phi>=|e^{(s_1)},g^{(s_2)}>$.
The upper curve  shows the results for $\gamma_1=0$ (Hermitian dynamics).
Then, from top to bottom, the curves for $\gamma_1=0.1, 0.5, 1$ are
displayed. All curves are drawn with statistical error bars
(which are already negligible by using just 50000 phase space points).
As expected the ``loss of probability'' (given by the fact that non-Hermitian
dynamics represents in an effective way the effect of additional states,
which do not appear in the Hamiltonian, whose occupation can grow at the
expense of the occupation of the explicitly described states; such is
the case when escaping from a well toward infinity or decaying from a
metastable state) increases upon increasing $\gamma_1$.
Figure~\ref{fig:fig2} displays the damped time evolution of the
reduced matrix element $\hat{\Omega}_{\rm S}^{22}$ of the spin chain.
The results confirm that the numerical algorithm is stable and that, at least
for the model considered, it can be used to simulate 
the decay of quantum states in a classical environment.

\begin{figure}
\resizebox{7cm}{5cm}{
\includegraphics* {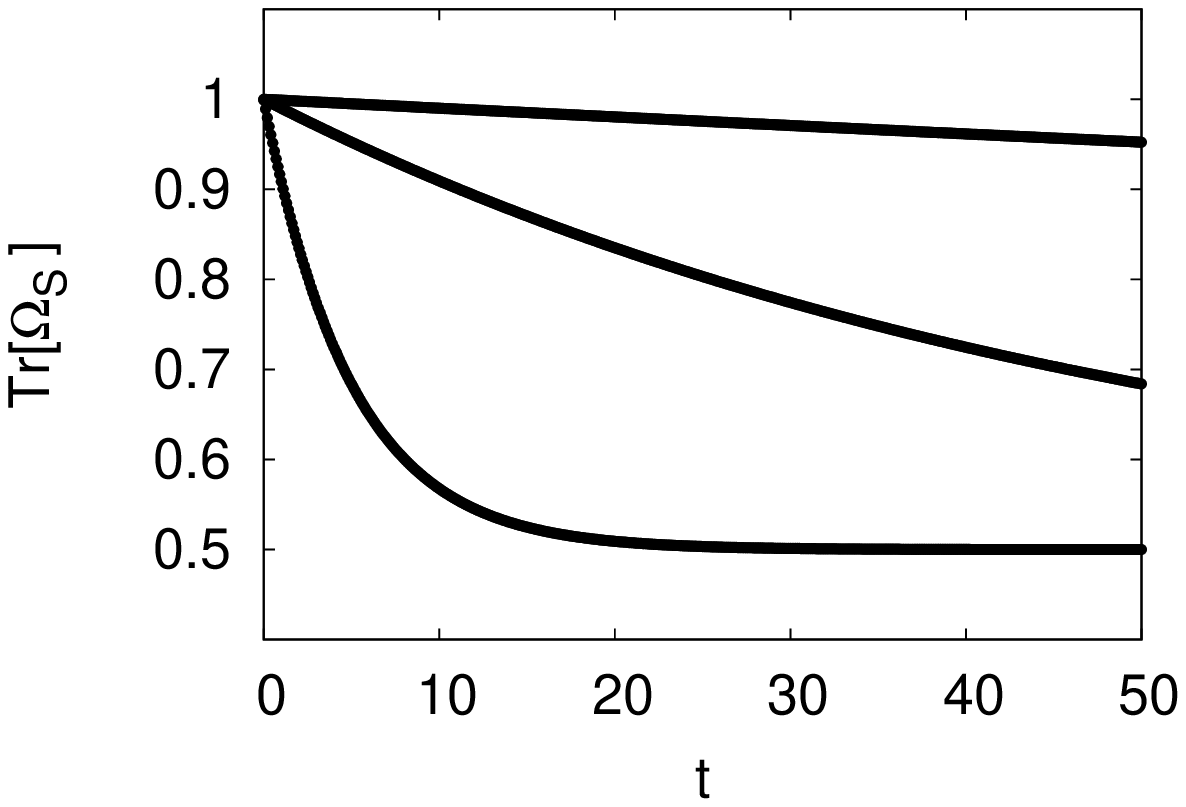}
}
\caption{
Adiabatic time evolution of the trace of the reduced density matrix,
${\rm Tr}[\Omega_{\rm S}]$,
for $\beta=0.1$, $j_x=j_y=-1$, $j_z=0.5$, $c=0.24$.
The numerical time step of integration is $\Delta t=0.01$.
Adimensional parameters are used.
Results for non-Hermitian dynamics with decay operator
$\hat{\Gamma}^{(2)}=\gamma_2|e^{(s_1)},e^{(s_2)}><e^{(s_1)},e^{(s_2)}|$
and initial reduced density matrix
$\hat{\Omega}_{\rm S}(t_0)=|\Psi><\Psi|$,
with
$|\Psi>=(1/\sqrt{2})\left(|e^{(s_1)},e^{(s_2)}>-|e^{(s_1)},g^{(s_2)}>\right)$.
The curves, from top to bottom,  show the results for
$\gamma_2=0.001,0.01,0.1$.
All curves are drawn with statistical error bars.
}
\label{fig:fig3}
\end{figure}

\begin{figure}
\resizebox{7cm}{5cm}{
\includegraphics* {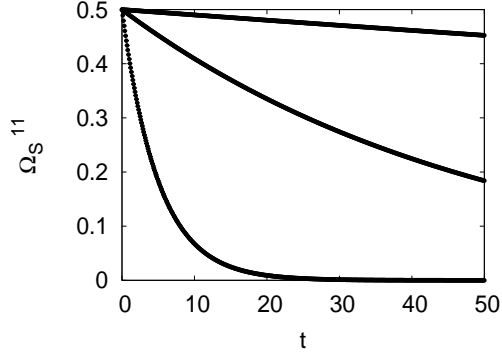}
}
\caption{
Adiabatic time evolution of matrix element $\Omega_{\rm S}^{22}$,
in the subsystem basis,
for $\beta=0.1$, $j_x=j_y=-1$, $j_z=0.5$, $c=0.24$.
The numerical time step of integration is $\Delta t=0.01$.
Adimensional parameters are used.
Results for non-Hermitian dynamics with decay operator
$\hat{\Gamma}^{(2)}=\gamma_2|e^{(s_1)},e^{(s_2)}><e^{(s_1)},e^{(s_2)}|$
and initial reduced density matrix
$\hat{\Omega}_{\rm S}(t_0)=|\Psi><\Psi|$,
with
$|\Psi>=(1/\sqrt{2})\left(|e^{(s_1)},e^{(s_2)}>-|e^{(s_1)},g^{(s_2)}>\right)$.
The curves, from top to bottom,  show the results for
$\gamma_2=0.001,0.01,0.1$.
All curves are drawn with statistical error bars.
}
\label{fig:fig4}
\end{figure}

Figure~\ref{fig:fig3} displays the
adiabatic time evolution of the trace of the reduced density matrix 
of the spin chain, ${\rm Tr}[\Omega_{S}]$,
when the decay operator is 
$\hat{\Gamma}^{(2)}=\gamma_2|e^{(s_1)},e^{(s_2)}><e^{(s_1)},e^{(s_2)}|$
and the initial reduced density matrix
$\hat{\Omega}_{\rm S}(0)=|\Psi><\Psi|$,
with
$|\Psi>=(1/\sqrt{2})\left(|e^{(s_1)},e^{(s_2)}>-|e^{(s_1)},g^{(s_2)}>\right)$.
The upper curve  shows the results for $\gamma_2=0.001$.
Then, from top to bottom, the curves for $\gamma_2=0.01$ and $\gamma_2=0.1$ are
shown.
Figure~\ref{fig:fig4} displays the corresponding damping phenomenon for the
reduced matrix element $\hat{\Omega}_{\rm S}^{11}$ of the spin chain.
In this case, the trace decays  because of the depletion of the state
$|e^{(s_1)},e^{(s_2)}>$, as it can be verified by monitoring the time
evolution of the diagonal elements of the density matrix in the basis
of the spin system.

\section{Conclusions}
\label{sec:conclusions}

In this work, a formalism to embed non-Hermitian quantum dynamics
in a classical bath has been provided.
In order to achieve this,
a quantum-classical approximation for the non-Hermitian equations of motion
of composite systems (with degrees of freedom having light
and heavy masses, $m$ and $M$, respectively) has been first been considered,
using a partial Wigner representation.
Then, the limiting case when the non-Hermitian part of the evolution
does not involve the classical-like degrees of freedom
has been taken into account.
The classical bath embedding the quantum system with non-conserved probability 
can be used as a noise source that is more general than those of
Gaussian type, \emph{e.g}, colored noise. The bath can be also
used to describe thermal disorder.

When the adiabatic part of the Hermitian Hamiltonian is considered, 
its eigenstates (defining the adiabatic basis)
can be used to represent the non-Hermitian quantum-classical equation of motion.
Once the equations of motion are represented in this adiabatic basis,
algorithms can be developed using a sequential short-time propagation scheme.
For the sake of illustrating the formalism, a Heisenberg chain with
two spins, each weakly coupled to a separate harmonic oscillator
has been studied.
Two different decay operators have been explicitly considered
showing that the algorithms lead to a stable and efficient numerical approach.

Future applications will be devoted to the modeling of nano-scale solid state
devices in dissipative environments.

\section*{Acknowledgements}

This work is based upon research supported by
the National Research Foundation of South Africa.

\appendix

\section{Representation of the equation of motion}
\label{app:derivation}

Starting from Eq.~(\ref{eq:Omega_W_ad_rep_1st})
the derivation can proceed by considering the term
\begin{eqnarray}
\langle\alpha;R|[\hat{\Gamma}_{\rm W},\hat{\Omega}_{\rm W}(X,t)
]_+|\alpha';R\rangle
&=&
\sum_{\beta\beta'}
\left(
\Gamma_{\rm W}^{\alpha\beta}\delta_{\alpha'\beta'}
+
\Gamma_{\rm W}^{\beta'\alpha'}\delta_{\alpha\beta}
\right)
\Omega_{\rm W}^{\beta\beta'}(X,t)
\;. \nonumber\\
\label{eq:ad_rep_comm}
\end{eqnarray}
Using Eq.~(\ref{eq:ad_rep_comm}), 
the equation of motion in~(\ref{eq:Omega_W_ad_rep_1st}) becomes
\begin{eqnarray}
\frac{\partial}{\partial t}\Omega_{\rm W}^{\alpha\alpha'}
=
&-&\sum_{\beta\beta'}i{\cal L}_{\alpha\alpha',\beta\beta'}
\Omega_{\rm W}^{\beta\beta'}
-\frac{1}{2\hbar}\sum_{\beta\beta'}
\left(\Gamma_{\rm W}^{\alpha\beta}\delta_{\alpha'\beta'}
+\Gamma_{\rm W}^{\beta'\alpha'}\delta_{\alpha\beta} \right)
\Omega_{\rm W}^{\beta\beta'}(X,t)
\nonumber\\
&-&\frac{i}{2}
\langle\alpha;R|\left\{\hat{\Gamma}_{\rm W},\hat{\Omega}_{\rm W}(X,t)\right\}
|\alpha';R\rangle
-\frac{i}{2}\langle\alpha;R|
\left\{\hat{\Omega}_{\rm W}(X,t),\hat{\Gamma}_{\rm W}\right\}
|\alpha';R\rangle
\;. \nonumber\\
\label{eq:Omega_W_ad_rep_2nd}
\end{eqnarray}
One can then consider the third term in the right hand side of
Eq.~(\ref{eq:Omega_W_ad_rep_2nd}):
\begin{eqnarray}
\langle\alpha;R|\left\{\hat{\Gamma}_{\rm W},\hat{\Omega}_{\rm W}(X,t)\right\}
|\alpha';R\rangle
&=&
\sum_{\gamma}
\left(
\langle\alpha;R|\frac{\partial\hat{\Gamma}_{\rm W}}{\partial R}
|\gamma;R\rangle
\frac{\partial\Omega_{\rm W}^{\gamma\alpha'}(X,t)}{\partial P}
\right.\nonumber\\
&-&\left.
\frac{\partial\Gamma_{\rm W}^{\alpha\gamma}}{\partial P}
\langle\gamma;R|
\frac{\partial\hat{\Omega}_{\rm W}(X,t)}{\partial R}
|\alpha';R\rangle
\right)
\;. 
\label{eq:GWOW_Poisson_aaprime}
\end{eqnarray}
Using the identities in Eqs.~(\ref{eq:iden1}-\ref{eq:iden2}),
Eq.~(\ref{eq:GWOW_Poisson_aaprime}) becomes
\begin{eqnarray}
\langle\alpha;R|\left\{\hat{\Gamma}_{\rm W},\hat{\Omega}_{\rm W}(X,t)\right\}
|\alpha';R\rangle
&=&
\sum_{\beta\beta'}
\left(
\frac{\partial\Gamma_{\rm W}^{\alpha\beta}}{\partial R}
\delta_{\alpha'\beta'}
\frac{\partial }{\partial P}
\right.\nonumber\\
&-& 
\left.
\frac{\partial\Gamma_{\rm W}^{\alpha\beta}}{\partial P}
\delta_{\alpha'\beta'}
\frac{\partial}{\partial R}\right)\Omega_{\rm W}^{\beta\beta'}(X,t)
\nonumber\\
&-&\sum_{\beta\beta'}
\left[\sum_{\sigma} \left(d_{\sigma\alpha}^* \Gamma_{\rm W}^{\sigma\beta}
\delta_{\alpha'\beta'} \frac{\partial}{\partial P}
\right.\right.\nonumber\\
&+&
\left.\left.
\Gamma_{\rm W}^{\alpha\sigma} d_{\sigma\beta}
\delta_{\alpha'\beta'}
\frac{\partial}{\partial P}
\right) \right]
\Omega_{\rm W}^{\beta\beta'}(X,t)
\nonumber\\
&+&
\sum_{\beta\beta'}
\left(
\sum_{\sigma}\frac{\partial\Gamma_{\rm W}^{\alpha\sigma}}{\partial P}
d_{\beta\sigma}^*\delta_{\alpha'\beta'}
\right.\nonumber\\
&+&
\left.
\frac{\partial\Gamma_{\rm W}^{\alpha\beta}}{\partial P}
d_{\beta'\alpha'}\right)
\Omega_{\rm W}^{\beta\beta'}(X,t)
\;.  \label{eq:GWOW_Poisson_aaprime_2nd}
\end{eqnarray}
Using Eq.~(\ref{eq:GWOW_Poisson_aaprime_2nd}),
Eq.~(\ref{eq:Omega_W_ad_rep_2nd}) becomes
\begin{eqnarray}
\frac{\partial}{\partial t}\Omega_{\rm W}^{\alpha\alpha'}
=
&-&\sum_{\beta\beta'}i{\cal L}_{\alpha\alpha',\beta\beta'}
\Omega_{\rm W}^{\beta\beta'}
-\frac{1}{\hbar}\sum_{\beta\beta'}
\left( \Gamma_{\rm W}^{\alpha\beta}\delta_{\alpha'\beta'}
+ \Gamma_{\rm W}^{\beta'\alpha'}\delta_{\alpha\beta} \right)
\Omega_{\rm W}^{\beta\beta'}(X,t)
\nonumber\\
&-&\frac{i}{2}
\sum_{\beta\beta'}
\left[
\left(
\frac{\partial\Gamma_{\rm W}^{\alpha\beta}}{\partial R}
\delta_{\alpha'\beta'}
\frac{\partial }{\partial P}
-
\frac{\partial\Gamma_{\rm W}^{\alpha\beta}}{\partial P}
\delta_{\alpha'\beta'}
\frac{\partial}{\partial R}\right)
\right.
\nonumber\\
&-&
\sum_{\sigma} \left(d_{\sigma\alpha}^* \Gamma_{\rm W}^{\sigma\beta}
\delta_{\alpha'\beta'} \frac{\partial}{\partial P}
+\Gamma_{\rm W}^{\alpha\sigma} d_{\sigma\beta}
\delta_{\alpha'\beta'}
\frac{\partial}{\partial P}
\right) 
\nonumber\\
&+&
\left.
\sum_{\sigma}\frac{\partial\Gamma_{\rm W}^{\alpha\sigma}}{\partial P}
d_{\beta\sigma}^*\delta_{\alpha'\beta'}
+
\frac{\partial\Gamma_{\rm W}^{\alpha\beta}}{\partial P}
d_{\beta'\alpha'}
\right]
\Omega_{\rm W}^{\beta\beta'}(X,t)
\nonumber\\
&-&\frac{i}{2}\langle\alpha;R|
\left\{\hat{\Omega}_{\rm W}(X,t),\hat{\Gamma}_{\rm W}\right\}
|\alpha';R\rangle\;.
\label{eq:Omega_W_ad_rep_3rd}
\end{eqnarray}
In order to complete the derivation, one must consider the last term in
the right hand side of Eq.~(\ref{eq:Omega_W_ad_rep_3rd}):
\begin{eqnarray}
\langle\alpha;R|
\left\{\hat{\Omega}_{\rm W}(X,t),\hat{\Gamma}_{\rm W}\right\}
|\alpha';R\rangle
&=&
\sum_{\gamma}
\langle\alpha;R|
\frac{\partial\hat{\Omega}_{\rm W}(X,t)}{\partial R}
|\gamma;R\rangle
\frac{\partial\hat{\Gamma}_{\rm W}^{\gamma\alpha'}}{\partial P}
\nonumber\\
&-&
\sum_{\gamma}
\frac{\partial\hat{\Omega}_{\rm W}(X,t)^{\alpha\gamma}}{\partial P}
\langle\gamma;R|
\frac{\partial\hat{\Gamma}_{\rm W}}{\partial R}
|\alpha';R\rangle
\;.  \label{eq:OWGW_Poisson_aaprime}
\end{eqnarray}
To further simplify Eq.~(\ref{eq:OWGW_Poisson_aaprime}),
one has to consider again (just with different indices) the identities given 
in Eqs.~(\ref{eq:iden1}-\ref{eq:iden2}).
Using Eqs.~(\ref{eq:iden1}-\ref{eq:iden2}), 
Eq.~(\ref{eq:OWGW_Poisson_aaprime}) provides
the final representation of the equation of motion
in the adiabatic basis.


\end{document}